\documentclass[aps,pra,showpacs]{revtex4}
\usepackage{amsmath}
\usepackage{graphicx}
\usepackage{dcolumn}
\usepackage{longtable}
\begin{document}

\title{Orbital precession due to central-force perturbations}

\author{Gregory S. Adkins}
\email[]{gadkins@fandm.edu}
\author{Jordan McDonnell}
\affiliation{Department of Physics, Franklin and Marshall College, Lancaster, Pennsylvania 17604}

\date{\today}

\begin{abstract}
We calculate the precession of Keplerian orbits under the influence of arbitrary central-force perturbations.  Our result is in the form of a one-dimensional integral that is straightforward to evaluate numerically.  We demonstrate the effectiveness of our formula for the case of the Yukawa potential.  We obtain analytic results for potentials of the form $V(r) = \alpha r^n$ and $V(r) = \alpha \ln(r/\lambda)$ in terms of the hypergeometric function ${_2F_1}\left ( \frac{1}{2}-\frac{n}{2},1-\frac{n}{2}; \, 2; \, e^2 \right )$, where $e$ is the eccentricity.  Our results reproduce the known general relativistic ($n=-3$), constant force ($n=1$), and cosmological constant ($n=2$) precession formulas.  Planetary precessions are often used to constrain the sizes of hypothetical new weak forces--our results allow for more precise, and often stronger, constraints on such proposed new forces.
\end{abstract}


\maketitle


\section{Introduction}

By the early years of the twentieth century, several workers had calculated the expected precession of the orbit of the planet Mercury based on the equations of Newtonian mechanics taking into account the perturbing effects of other planets in the solar system.  (See Doolittle \cite{Doolittle12} for a discussion of methods, a summary of results, and early references, and Stewart \cite{Stewart05} for a recent discussion.)  The prediction was for a precession of $575$ arcsec per century (in a sun-centered coordinate system).  Observation found the actual precession to be $532$ arcsec per century, leaving $43$ arcsec per century unexplained.  This discrepancy remained a mystery until 1915, when Einstein, in the course of the development of general relativity, calculated the weak-field expansion of his new theory and predicted an additional precession of just this amount. \cite{Einstein15,Pais82}  Since then, the precession of Mercury and the other planets has been used to place bounds on the size of additional hypothetical forces.  Examples of such forces include the ``anti-gravity'' due to the cosmological constant 
\cite{Islam83,Iorio06a,Sereno06}, forces associated with a distance dependence of the gravitational constant \cite{Mikkelsen77}, forces proposed as alternatives to dark matter \cite{Gron96,Khriplovich06}, forces proposed as phenomenological explanations of the Pioneer Anomaly 
\cite{Anderson98,Anderson02,Jaekel06,Brownstein06,Iorio06b}, and forces induced from higher dimensional models \cite{Dvali00,Dvali03,Capozziello03,Gruzinov05} or extensions of general relativity 
\cite{Capozziello01,Moffat06,Sanders06}.

In most of the precession calculations, the assumption of nearly-circular orbital motion has been made in order to simplify the calculations.  We show in this work that such an approximation is not necessary, and give a simple formula for the calculation of the precession induced by any central force even for highly eccentric orbits.  Our result can be used to get numerical values for the precession in all cases, and analytic values for power-law and logarithmic potentials.  Since the exact precession values can differ from the nearly-circular values by a factor of two or more, significantly more precise results for constraints on proposed new forces can be obtained through use of the present work.

The outline of this work is as follows.  In Sec.~II we review the case of nearly-circular orbits and give a derivation of the relevant precession formula.  In Sec.~III we develop a formula for the precession of non-circular orbits in the form of a one-dimensional integral.  In Sec.~IV we apply our formula to obtain precession results for a perturbing force of Yukawa form as a function of the Yukawa range parameter and the orbital eccentricity.  In Sec.~V we obtain analytic results for power-law and logarithmic potentials.  Finally, in Sec.~VI we discuss a number of examples of hypothetical central-force perturbations for which precession bounds are appropriate.


\section{Precession in nearly-circular orbits}

We begin with a brief review of the Keplerian case.  The unperturbed orbit equation is
\begin{equation}
\frac{d^2 u}{d \varphi^2} + u = \frac{G M}{h^2} \, ,
\end{equation}
where $r=1/u$ and $\varphi$ are cylindrical coordinates in the orbital plane, $G$ is the gravitational constant, $M$ is the central mass, and $h = \dot \varphi/u^2$ is the angular momentum per unit mass of the orbiting body.  The standard elliptical solution has the form
\begin{equation} \label{unperturbed}
u_0(\varphi) = \frac{1}{L} \left \{ 1+e \cos(\varphi - \varphi_0 ) \right \} 
\end{equation}
where $L = h^2/(G M)$ is the semilatus rectum of the orbital ellipse and $e$ is its eccentricity.  The maximum and minimum distances from the orbiting body to the force center have the values
\begin{equation}
r_\pm = \frac{L}{1 \mp e} \, .
\end{equation}
The semimajor axis $a$ is thus related to the semilatus rectum according to
\begin{equation}
a = \frac{1}{2} (r_+ + r_-) = \frac{L}{1-e^2} \, ,
\end{equation}
so that $r_\pm = (1 \pm e) a$.  The period $\tau$ is given by Kepler's Third Law: $\tau^2 = 4 \pi^2 a^3/(G M)$.

In the limit of nearly circular orbits we can find an approximate solution to the orbital equation that displays the precession of the axes of the ellipse.  In this approach we generalize the method of Ohanian \cite{Ohanian76}.  We assume the presence of a radial perturbing force $F(r)$ in addition to the usual Newtonian gravitational force $-G M m/r^2$.  The perturbed orbit equation takes the form
\begin{equation}
\frac{d^2 u}{d \varphi^2} + u = \frac{G M}{h^2} - \frac{g(u)}{h^2}
\end{equation}
where $g(u) = r^2 \frac{F(r)}{m} \vert_{r=1/u}$.  The unperturbed solution is given in (\ref{unperturbed}) above.  We assume that the eccentricity $e$ is small, and expand
\begin{equation}
g(u) = g_0 + g_1 \frac{G M}{h^2} e \cos(\varphi - \varphi_0) + O(e^2)
\end{equation}
where $g_0 = g(1/L)$ and $g_1 = g'(1/L)$.  To first order in $e$ the orbit equation can be written as
\begin{equation}
\frac{d^2 u}{d \varphi^2} + u \approx \frac{G M - g_0}{h^2} - g_1 \frac{G M}{h^4} e \cos(\varphi - \varphi_0) \, .
\end{equation}
We try a solution
\begin{equation}
u(\varphi) = \left ( \frac{G M - g_0}{h^2} \right ) \left \{ 1 + e \cos \left [ (1+B) (\varphi - \varphi_0) \right ] \right \}
\end{equation}
and find, to first order in the perturbation, that $B = g_1/(2 h^2)$.  It follows that the precession angle (per revolution) is $\Delta \theta_p = - 2 \pi B = - \pi g_1/h^2$.  We notice immediately that a $1/r^2$ perturbation, having $g(u)=\text{const.}$, gives no precession.

If we write the perturbing force in terms of a potential
\begin{equation}
F(r) = -\frac{dV}{dr} \, ,
\end{equation}
we find that
\begin{equation}
g_1 = \frac{d}{du} \left ( -\frac{1}{m} r^2 \frac{dV}{dr} \right ) \vert_{u=1/L}
= \frac{1}{m} \frac{d^2 V}{du^2} \vert_{u=1/L}
\end{equation}
where we have expressed the potential $V(r)$ in terms of $u=1/r$.  It follows that the precession 
$\Delta \theta_p = -\pi g_1/h^2$ can be written in terms of the potential as
\begin{equation} \label{near_circ_result}
\Delta \theta_p = -\frac{\pi}{G M m} \frac{1}{L} \frac{d^2 V}{du^2} \vert_{u=1/L}
\end{equation}
where we have used $h^2=G M L$.  This is equivalent to a formula for the nearly-circular precession that has been used by Dvali, Gruzinov, and Zaldarriaga \cite{Dvali03}.

For a power-law perturbing potential $V(r) = \alpha_n r^n$ the results are particularly simple.  The force has the form $F(r) = - n \alpha_n r^{n-1}$, and the precession (per revolution) is
\begin{equation}
\Delta \theta_p(n) = -n (n+1) \pi \, \frac{\alpha_n L^{n+1}}{G M m} \, .
\end{equation}
There is an interesting relationship between the precession angles for powers $n$ and $-(n+1)$.  The precession when $n \rightarrow -(n+1)$ is
\begin{equation}
\Delta \theta_p(-(n+1)) = - n (n+1) \pi \, \frac{\alpha_{-(n+1)} L^{-n}}{G M m} \, ,
\end{equation}
so that
\begin{equation} \label{circular_power_relation}
\frac{\Delta \theta_p(n)}{\alpha_n} = L^{2n+1} \frac{\Delta \theta_p(-(n+1))}{\alpha_{-(n+1)}} \, .
\end{equation}


\section{Precession in non-circular orbits} 

In this section we will obtain an expression for the precession valid for orbits of any eccentricity and explore some of its properties.  In the following sections we will discuss the Yukawa force as an example where numerical techniques are required, and we will find analytic results for a number of useful potentials.

We will use energy conservation to  obtain a simple expression for the precession.  The conserved energy for motion in potential $-G M m/r+V(r)$ has the form
\begin{equation}
E = \frac{m}{2} \left ( \dot r^2 + \frac{h^2}{r^2} \right ) - \frac{G M m}{r} + V(r) \, .
\end{equation}
In terms of the more useful variable $u=1/r$ this is
\begin{equation}
E = \frac{m h^2}{2} \left ( \left ( \frac{du}{d \varphi} \right ) + u^2 \right ) - G M m u + V(u) \, .
\end{equation}
We solve for $du/d\varphi$ and integrate to obtain
\begin{equation}
\varphi - \varphi_0 = \pm \int \frac{du}{\left ( \frac{2E}{m h^2} + \frac{2 G M u}{h^2} - u^2 -
\frac{2 V(u)}{m h^2} \right )^{1/2}} \, ,
\end{equation}
where the sign is determined by the position in the orbit (increasing radius or decreasing radius).

Returning to the non-perturbed problem for a moment, we see that the turning points are the solutions $a_\pm=1/r_\mp = (1\pm e)/L$ to the equation
\begin{equation}
0 = \frac{2E}{m h^2} + \frac{2 G M u}{h^2} - u^2 = (a_+-u)(u-a_-) \, .
\end{equation}
The angle change for a full orbit (twice the change for the interval $u=a_-$ to $u=a_+$) is
\begin{equation}
\Delta \varphi = 2 \int_{a_-}^{a_+} \frac{du}{\sqrt{(a_+-u)(u-a_-)}}  = 2 \pi \, ,
\end{equation}
as expected.  Explicit expression for $a_\pm$ are not needed in the evaluation of this integral.

We are now prepared to write down the general expression for the angle increase per complete radial oscillation ($u$ from smallest to largest and back to smallest).  It is
\begin{equation}
\Delta \varphi = 2 \int_{u_-}^{u_+} \frac{du}{\left ( \frac{2E}{m h^2} + \frac{2 G M u}{h^2} - u^2 -
\frac{2 V(u)}{m h^2} \right )^{1/2}} \, ,
\end{equation}
where $u_\pm$ are the zeros of
\begin{equation}
D(u) \equiv \frac{2E}{m h^2} + \frac{2 G M u}{h^2} - u^2 - \frac{2 V(u)}{m h^2}
\end{equation}
with $u_\pm \approx a_\pm$.  The precession per revolution is thus 
$\Delta \theta_p = \Delta \varphi - 2 \pi$.

We will need approximate values for the zeros $u_\pm$.  We note that the expression $D(u)$ can be written as
\begin{equation} \label{first_form_D}
D(u) = (a_+-u)(u-a_-) - \frac{2 V(u)}{m h^2} \, .
\end{equation}
We write $u_\pm = a_\pm + \delta_\pm$, and find that the first-order shifts in the positions of the zeros are given by
\begin{equation}
\delta_\pm = \frac{\mp 2 V(a_\pm)}{(a_+-a_-) m h^2} \, .
\end{equation}

It is useful to express $D(u)$ with its zeros explicitly displayed as
\begin{equation}
D(u) = (u_+-u)(u-u_-) \left ( 1 + \frac{2 \cal{G}}{m h^2} \right ) \, .
\end{equation}
It is clear that $\cal{G}$ is small, of the order of the perturbation, since $D(u)$ can also be written as in (\ref{first_form_D}) and $u_\pm \approx a_\pm$.  An explicit approximation for $\cal{G}$ can be found by equating the two forms for $D(u)$.  We find that
\begin{equation}
{\cal G} \approx \frac{-(a_+-a_-) V(u) + (a_+-u) V(a_-) + (u-a_-) V(a_+)}{(a_+-a_-) 
(a_+-u)(u-a_-)} \, .
\end{equation}
We note that the numerator of ${\cal G}$ vanishes when $a_+ = a_-$ and when $u = a_\pm$, so ${\cal G}$ is non-singular for the whole physical range $a_- \le u \le a_+$.  With this expression for $D(u)$, the precession can be written as
\begin{eqnarray}
\Delta \theta_p &=& 2 \int_{u_-}^{u_+} \frac{du}{\sqrt{(u_+-u)(u-u_-)}} \left ( 1 + \frac{2 {\cal G}}{m h^2} \right )^{-1/2} - 2 \pi \nonumber \\
&\approx& - \frac{2}{m h^2} \int_{a_-}^{a_+} \frac{du \, {\cal G}}{\sqrt{(a_+-u)(u-a_-)}} \, .
\label{general_result_1}
\end{eqnarray}

A number of consistency checks can be applied to this result.  First, we notice that ${\cal G}$ vanishes when either $V(u) = \text{const.}$ (no force) or $V(u) \propto u$ ( a contribution to the $1/r^2$ force), so there is no precession in these cases.  Also, we can work out the precession for a nearly-circular orbit as a limiting case of (\ref{general_result_1}) for $a_+ \rightarrow a_-$.  When $a_+$, $a_-$, and $u$ are all similar, we can use  a series expansion to show that 
${\cal G} \rightarrow (1/2) V''(u)$.  The corresponding precession, from (\ref{general_result_1}), agrees with our earlier result (\ref{near_circ_result}).  For a power-law perturbing potential 
$V(r) = \alpha_n r^n$, the change of variable $u = a_+ a_-/x$ relates $\Delta \theta_p(n)$ and
$\Delta \theta_p(-(n+1))$.  We find that
\begin{equation} \label{elliptical_power_relation}
\frac{\Delta \theta_p(n)}{\alpha_n} = b^{2n+1} \frac{\Delta \theta_p(-(n+1))}{\alpha_{-(n+1)}} \, ,
\end{equation}
which is the generalization of (\ref{circular_power_relation}) to non-circular orbits.  Here 
$b = a \sqrt{1-e^2} = L/\sqrt{1-e^2} = 1/\sqrt{a_+ a_-}$ is the semiminor axis.

Occasionally it is more convenient to express the perturbation as a force instead of a potential, for instance when the phenomenological perturbing force has a complicated structure or is given as an anomalous acceleration.  Our precession result (\ref{general_result_1}) can be rewritten in terms of the force by means of an integration by parts (using $\int du ((a_+-u)(u-a_-))^{-3/2} = 2(2u-a_+-a_-) 
(a_+-a_-)^{-2} ((a_+-u)(u-a_-))^{-1/2}$) in the form
\begin{equation}
\Delta \theta_p = \frac{-4}{G M m L (a_+-a_-)^2} \int_{a_-}^{a_+} \frac{du (2u-a_+-a_-) V'(u)}
{\sqrt{(a_+-u)(u-a_-)}} \, .
\end{equation}
An additional advantage of this form is the absence of the potentially troublesome factor 
$((a_+-u)(u-a_-))^{3/2}$ in the denominator.  For practical use it is convenient to use a dimensionless integration variable with a fixed range.  We define $u=(1+e z)/L$, and find
\begin{equation} \label{general_result_2}
\Delta \theta_p = \frac{-2 L^2}{G M e} \int_{-1}^1 \frac{dz \, z}{\sqrt{1-z^2}} \frac{F(z)/m}{(1+e z)^2} 
\end{equation}
where $F(z)$ is the perturbing force at radius $r=L/(1+e z)$.  An alternative expression for the precession, in terms of the potential, is
\begin{equation} \label{general_result_3}
\Delta \theta_p = \frac{-2 L}{G M m e^2} \int_{-1}^1 \frac{dz \, z}{\sqrt{1-z^2}} \frac{d V(z)}{dz}
\end{equation}
where $V(z)$ is the perturbing potential at radius $r=L/(1+e z)$.

A general consequence of either form (\ref{general_result_2}) or (\ref{general_result_3}) is the fact that the precession $\Delta \theta_p$ contains only even powers of the eccentricity $e$.  This is true because only even powers of $z$ contribute in the integrals, and $e$ and $z$ are tied together in the definition of $r$.

Of course, celestial dynamics is an exceedingly well-developed discipline, and expressions equivalent to (\ref{general_result_2}) and (\ref{general_result_3}) can be derived from the formulas of orbital perturbation theory given in the standard texts (see for example Danby \cite{Danby62}).  Landau and Lifshitz \cite{Landau76} give a precession result in an alternative, but also equivalent, form.  We have found the expressions (\ref{general_result_2}) and (\ref{general_result_3}) useful for further developments.


\section{The Yukawa force}

The Yukawa potential
\begin{equation}
V(r) = \alpha \frac{e^{-r/\lambda}}{r}
\end{equation}
of strength $\alpha$ and range $\lambda$ is a field-theory-motivated \cite{Yukawa35,Nieto91} assumption for a correction to the gravitational force.  In this section, we work out the precession due to a perturbation of Yukawa form.

The precession $\Delta \theta_p(\kappa,e)$ due to a Yukawa perturbation depends on two parameters: a range parameter $\kappa=L/\lambda$ and the eccentricity.  It seems difficult to obtain an analytic form for $\Delta \theta_p(\kappa,e)$, so we work numerically instead.  Use of either (\ref{general_result_2}) or (\ref{general_result_3}) leads to the integral form
\begin{equation}
\Delta \theta_p(\kappa,e) = -\frac{2 \alpha}{G M m e} \int_{-1}^1 \frac{dz \, z}{\sqrt{1-z^2}} 
\left ( 1+ \frac{\kappa}{1+e z} \right ) \exp \left ( {\frac{-\kappa}{1+e z}} \right ) \, .
\end{equation}
\begin{figure}
\includegraphics[width=4.0in]{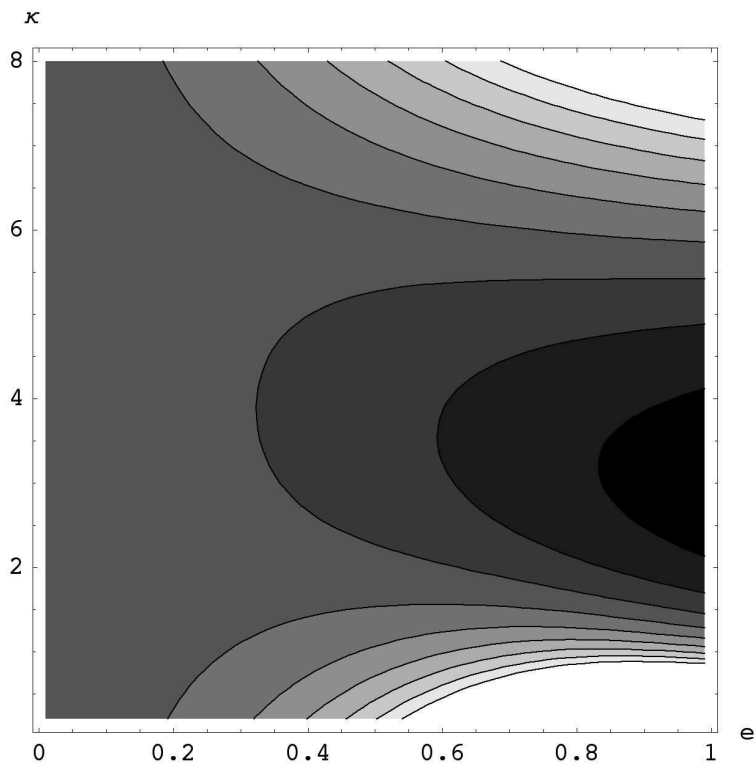}
\caption{Contour plot showing the relative precession $I(\kappa,e)$ for the Yukawa potential as a function of the range parameter $\kappa$ and eccentricity $e$.  The contour lines are at $0.75$, $0.85$, $\dots$, $1.55$.  The darkest (lightest) shading corresponds to the smallest (largest) values for the relative precession.  The medium gray shown on the left side of the plot corresponds to values near one.  As expected, the largest variations from one occur for large eccentricity.  Surprisingly, there is a range of $\kappa$ for which the relative precession is depressed from its $e=0$ value.}
\label{figure1} 
\end{figure}
The small-$e$ limit of this, obtained either as the first non-vanishing term in an expansion in $e$ or by use of (\ref{near_circ_result}) directly, is
\begin{equation}
\Delta \theta_p(\kappa,0) = \frac{- \pi \alpha}{G M m} \kappa^2 e^{-\kappa} \, .
\end{equation}
We take $\Delta \theta_p(\kappa,0)$ to set the scale for the Yukawa precession, and focus instead on the effect of eccentricity encoded in the relative precession $I(\kappa,e)$ where
\begin{equation}
\Delta \theta_p(\kappa,e) = \Delta \theta_p(\kappa,0) I(\kappa,e) \, .
\end{equation}
The integral for the relative precession is
\begin{equation}
I(\kappa,e) = \frac{2}{\pi \kappa^2 e} \int_{-1}^1 \frac{dz \, z}{\sqrt{1-z^2}} 
\left ( 1+ \frac{\kappa}{1+e z} \right ) \exp \left ( {\frac{\kappa e z}{1+e z}} \right ) \, .
\end{equation}
The relative precession can give an important correction to the zero-eccentricity result.  For example, the relative precession of the asteroid Icarus (with eccentricity e=0.827) with a Yukawa perturbation having range parameter $\kappa = 0.1$ would be $I(0.1,0.827)=4.57$.  The general behavior of $I(\kappa,e)$ as a function of its parameters is shown in Fig.~1.  For most values of its parameters $I(\kappa,e)>1$, and the constraints on the Yukawa strength parameter $\alpha$ from precession bounds are strengthened by taking the eccentricity dependence of $I(\kappa,e)$ into account.


\section{Analytic results for power-law and logarithmic potentials}

For some perturbing potentials it is possible to obtain analytic precession results.  In this section we work out the analytic results for power law and logarithmic perturbations.  There are known expressions for the precession due to a few particular power law perturbations.  The small-field expansion of general relativity, valid outside a spherically symmetric mass, leads to a potential proportional to $r^{-3}$.  A constant force has a potential proportional to $r$.  Also, a uniform dark energy force is modeled by a potential proportional to $r^2$.  We give a general expression for a perturbative potential $V(r) = \alpha_n r^n$ that contains these results as special cases.

For a power law potential we take $V(r) = \alpha_n r^n$, and apply (\ref{general_result_3}).  It turns out to be easiest to work first with a negative exponent.  In particular, for a potential 
$V(r) = \alpha_{-(n+1)} r^{-(n+1)} = \alpha_{-(n+1)} (1+e z)^{n+1}/L^{n+1}$, the precession is
\begin{equation}
\Delta \theta_p(-(n+1)) = \frac{-2 \alpha_{-(n+1)} (n+1)}{G M m L^n e} \int_{-1}^1 \frac{dz \, z}{\sqrt{1-z^2}}
(1+e z)^n \, .
\end{equation}
The integral can be done by expanding $(1+e z)^n$ in a series and integrating term by term.  One finds that
\begin{eqnarray} \label{integral_formula}
\frac{1}{e} \int_{-1}^1 \frac{dz \, z}{\sqrt{1-z^2}} (1+e z)^n
&=& \frac{1}{e} \sum_{i=0}^\infty \begin{pmatrix} n \\ 2 i + 1 \end{pmatrix} 2 \int_0^1
\frac{dz \, z}{\sqrt{1-z^2}} (e z)^{2i+1} \nonumber \\
&=& \sum_{i=0}^\infty e^{2i} \begin{pmatrix} n \\ 2 i + 1 \end{pmatrix} \pi
\frac{(2i+1)!!}{(2i+2)!!} \nonumber \\
&=& \frac{\pi n}{2} \, {_2F_1} \left (\frac{1}{2}-\frac{n}{2},1-\frac{n}{2}\, ;2\, ;e^2 \right ) \, .
\end{eqnarray}
In the first step here we made use of the fact that only even powers of $z$ survive the integration.  Our complete result then for negative powers is
\begin{equation} \label{power_law_negative}
\Delta \theta_p(-(n+1)) = \frac{-\pi \alpha_{-(n+1)}}{G M m} \frac{\chi_n(e)}{L^n}
\end{equation}
where
\begin{equation}
\chi_n(e) = n (n+1) \, {_2F_1} \left (\frac{1}{2}-\frac{n}{2},1-\frac{n}{2}\, ;2\, ;e^2 \right ) \, .
\end{equation}
The precession for positive powers can be obtained by using (\ref{elliptical_power_relation}) or the hypergeometric identity \cite{Gradshteyn80a}
\begin{equation}
\chi_n(e) = (1-e^2)^{n+1/2} \chi_{-(n+1)}(e) \, ,
\end{equation}
and is
\begin{equation} \label{power_law_positive}
\Delta \theta_p(n) = \frac{-\pi \alpha_n}{G M m} a^{n+1} \sqrt{1-e^2} \chi_n(e) \, .
\end{equation}
We have chosen to write $\Delta \theta_p$ in terms of the semilatus rectum $L$ for negative powers and in terms of the semimajor axis $a$ for positive powers in order to simplify the formulas.  We note that expressions (\ref{power_law_negative}) and (\ref{power_law_positive}) are in fact both valid for all $n$.

The hypergeometric function ${_2F_1}$ terminates whenever one of its first two arguments is zero or a negative integer, which happens for any natural number $n$.  More generally, ${_2F_1}$ and 
$\Delta \theta_p(n)$ are well defined for any $n$, integral or not.  Results for the first few polynomials $\chi_n(e)$ are given in Table~I.

\begin{table}[t]
\caption{\label{table1} Polynomials $\chi_n(e)$ for various values of $n$.}
\begin{tabular}{cc}
$n$ & $\chi_n(e)$ \\
\hline\noalign{\smallskip}
$1 \;$ & $2$ \\
$2 \;$ & $6$ \\
$3 \;$ & $3 (4 + e^2)$ \\
$4 \;$ & $5 (4 + 3 e^2)$ \\
$5 \;$ & $\frac{15}{4} (8 + 12 e^2 + e^4)$ \\
$6 \;$ & $\frac{21}{4} (8 + 20 e^2 + 5 e^4)$ \\
$7 \;$ & $\frac{7}{8} (64 + 240 e^2 + 120 e^4 + 5 e^6)$ \\
\noalign{\smallskip}\hline
\end{tabular}
\end{table}

Our power law expressions can be put to immediate use to reproduce known results for $n=-3$, $n=1$, and $n=2$.  Einstein \cite{Einstein15} considered orbits around a spherically symmetric mass in the weak-field limit.  The first post-Newtonian correction is a perturbing potential $V(r) = - \frac{G M m h^2}{c^2 r^3}$ where $h^2=G M L$.  The corresponding precession from 
(\ref{power_law_negative}) is
\begin{equation} \label{gr_precession}
\Delta \theta_p(\text{GR}) = \frac{6 \pi G M}{c^2 L} \, .
\end{equation}
This gives the famous $43$ arcsec per century when applied to the orbit of Mercury.  A constant perturbing radial force $F=m a_0$ is represented by the potential $V(r) = -m a_0 r$.  The associated precession (per revolution) from (\ref{power_law_positive}) is
\begin{equation}
\Delta \theta_p(F=m a_0) = \frac{2 \pi a_0 a^2}{G M} \sqrt{1-e^2} \, ,
\end{equation}
consistent with the result of Sanders \cite{Sanders06}.
The extra-Newtonian radial force due to a uniform dark energy is given by \cite{Rindler06}
\begin{equation}
F(r) = \frac{1}{3} m \Lambda c^2 r \, .
\end{equation}
This corresponds to a perturbing potential $V(r) = -\frac{1}{6} m \Lambda c^2 r^2$.   The corresponding dark energy induced precession (per revolution) from (\ref{power_law_positive}) is
\begin{equation} \label{de_precession}
\Delta \theta_p(\Lambda) = \frac{\pi \Lambda c^2 a^3}{G M} \sqrt{1-e^2}
\end{equation}
in agreement with the expression of Kerr, Hauck, and Mashhoon \cite{Kerr03}.  (We note the curious fact that the $n=-3$ general relativity precession (\ref{gr_precession}) and the $n=2$ dark energy precession (\ref{de_precession}) are dual to each other in the sense of (\ref{elliptical_power_relation}).)  Result (\ref{de_precession}) can be used to find a bound on the cosmological constant $\Lambda$.  The measured anomalous precession for Mercury, after accounting for known Newtonian and relativistic effects and solar oblateness, is 
$-0.0036(50)$ arcsec per century \cite{Pitjeva05}.  The orbital parameters for Mercury are
$a = 5.79 \times 10^{10}m$, $e=0.206$, $\tau = 0.241y = 7.60 \times 10^6 s$, and the solar mass is
$M = M_\odot = 1.99 \times 10^{30} kg$, leading to a bound on $\Lambda$ of
\begin{equation}
-2.5 \times 10^{-44} cm^{-2} < \Lambda < 0.4 \times 10^{-44} cm^{-2} \, .
\end{equation}
Tighter bounds on $\Lambda$ can be found by using precession data for other planets. \cite{Iorio06a}  These bounds are several orders of magnitude larger than the currently accepted value of 
$\Lambda$ found from cosmological and large-scale measurements, but it is interesting to find bounds on the cosmological constant based on purely solar system measurements \cite{Islam83}.

An analytic result can also be obtained for a logarithmic potential.  We assume a perturbative potential modification of the form
\begin{equation}
V(r) = \alpha \ln \left ( \frac{r}{\lambda} \right )
\end{equation}
where $\lambda$ is an arbitrary scale parameter.  Then $V'(z) = -\alpha e/(1+e z)$, and from (\ref{general_result_3}) the associated precession is
\begin{equation}
\Delta \theta_p(\text{log}) = \frac{2 \alpha L}{G M m e} \int_{-1}^1 \frac{dz \, z}{\sqrt{1-z^2}} (1+e z)^{-1} \, .
\end{equation}
We have already done the required integral in (\ref{integral_formula}):
\begin{equation}
\frac{1}{e} \int_{-1}^1 \frac{dz \, z}{\sqrt{1-z^2}} (1+e z)^{-1}
= - \frac{\pi}{2} {_2 F_1} \left ( 1, \frac{3}{2}; 2; e^2 \right ) = - \frac{\pi}{e^2} \left ( \frac{1}{\sqrt{1-e^2}} -1 \right ) \, ,
\end{equation}
where \cite{Gradshteyn80b} was used to obtain the value of the hypergeometric function.  The precession for a logarithmic perturbation is
\begin{equation} \label{log_potential}
\Delta \theta_p(\text{log}) = \frac{- 2 \pi \alpha L}{G M m e^2} \left ( \frac{1}{\sqrt{1-e^2}} -1 \right ) \, .
\end{equation}
We note that $\Delta \theta_p(\text{log}) = \lim_{n \rightarrow 0} \frac{\Delta \theta_p(n)}{n}$.


\section{Summary and Discussion}

We have found expressions for the precession induced by central-force perturbations that generalize the known nearly-circular result (\ref{near_circ_result}).  The most general results are 
(\ref{general_result_2}) for perturbing force $F(z)$, and (\ref{general_result_3}) for perturbing potential $V(z)$.  These expressions are straightforward to evaluate numerically.  For power-law and logarithmic perturbing potentials, simple analytic results can be found: (\ref{power_law_negative}) and 
(\ref{power_law_positive}) for power-law potentials, and (\ref{log_potential}) for a logarithmic potential.
As examples of the use of these formulas, we have shown in some detail how the precession for the Yukawa potential can be found numerically.  Also, we applied our formulas to reproduce the known precession results for the general relativistic $n=-3$ potential, the constant force $n=1$ potential, and the cosmological constant $n=2$ potential.

Many other central-force modifications to gravity can be found in the literature, all of which can be treated according to the methods developed here.  A small sample of proposed modifications to gravity that have been discussed in the recent literature include the following.

There are many proposals for modifications to gravity in our four-dimensional world that arise from a more fundamental theory in a higher dimension.  An example is the five-dimensional braneworld scenario of Dvali, Gabadadze, and Porrati \cite{Dvali00}, in which the short distance limit of four-dimensional gravity has the form
\begin{equation}
V(r) = \frac{- G M m}{r} \left \{ 1+ \frac{2}{\pi} \left [ \ln \left ( \frac{r}{r_0} \right ) + \gamma -1 \right ] \left ( \frac{r}{r_0} \right ) + O(r^2) \right \} \, .
\end{equation}
Further study of this model by Gruzinov \cite{Gruzinov05} and by Dvali, Gruzinov, and Zaldarriaga  
\cite{Dvali03} points to a perturbing potential proportional to $r^{1/2}$.  Capozziello \textit{et al.} \cite{Capozziello01} discuss the effect of general Yukawa corrections 
\begin{equation}
V(r) = -\frac{G M m}{r} \left \{ 1 + \sum_{k=1}^n \alpha_k e^{-r/\lambda_k} \right \} \, .
\end{equation}
Capozziello and Lambiase \cite{Capozziello03} studied the Newtonian limit of string-dilation gravity, and found a perturbing potential containing a linear combination of Yukawa, $r^2$, and $\cosh(r/\lambda)$ terms.

The Pioneer Anomaly, the anomalous acceleration of about $8.5 \times 10^{-10} m/s^2$ for $20 AU \le r \le 70 AU$ directed towards the sun experienced by the Pioneer 10/11 spacecraft, has been a strong source of motivation for the development of modified theories of gravity.  One model of the proposed gravitational force considered by Anderson \textit{et al.} \cite{Anderson98,Anderson02} is a Yukawa potential (Newtonian plus perturbation) that has the usual Newtonian $1/r^2$ force but no $1/r$ correction:
\begin{equation}
V(r) = - \frac{G M m}{r} \frac{1+\alpha e^{-r/\lambda}}{1+\alpha} \, .
\end{equation}
Jaekel and Reynaud \cite{Jaekel06} consider models involving linear and quadratic perturbing potentials.  Brownstein and Moffat \cite{Brownstein06} discuss the Newtonian limit of the scalar-tensor vector gravity theory \cite{Moffat06}, which is best expressed as an acceleration of Yukawa-like form but with scale-dependent parameters:
\begin{equation}
\frac{F(r)}{m} = - \frac{G M}{r^2} \left \{ 1 + \alpha(r) \left [ 1-e^{-r/\lambda(r)} \left ( 1+ \frac{r}{\lambda(r)} \right ) \right ] \right \} \, .
\end{equation}
For any explicitly given strength and range parameters $\alpha(r)$ and $\lambda(r)$, the precession due to such an anomalous acceleration can be evaluated numerically using (\ref{general_result_2}).
Sanders \cite{Sanders06} discussed solar-system constraints on multi-field extensions to general relativity that might explain the Pioneer Anomaly.

For order-of-magnitude estimates of orbital precession it is often adequate to use the near-circular formula (\ref{near_circ_result}).  In some situations, especially for highly eccentric orbits, and whenever precision is desired, precessions should be calculated taking into account the eccentricity of the orbit.  This is straightforward to do numerically for any perturbing potential or force, and for power-law and logarithmic perturbing potentials analytic results are now available.


\begin{acknowledgments}
We are grateful to our colleagues Fronefield Crawford, Richard Fell, Christie Larochelle, Michael McCooey, and Calvin Stubbins for useful comments and advice.  We acknowledge the financial support of the Hackman Scholars and Marshall Scholars programs of Franklin \& Marshall College.
\end{acknowledgments}


\end{document}